# Collisional Charging of Individual Sub-Millimeter Particles: Using Ultrasonic Levitation to Initiate and Track Charge Transfer

Victor Lee,[1,2] Nicole M. James,[1,3] Scott Waitukaitis,[4,5] and Heinrich M. Jaeger[1,2]

[1]James Franck Institute, University of Chicago, Chicago, IL 60637, USA
[2]Department of Physics, University of Chicago, Chicago, IL 60637, USA
[3]Department of Chemistry, University of Chicago, Chicago, IL 60637, USA
[4]AMOLF, 1098 XG Amsterdam, The Netherlands
[5]Huygens-Kamerlingh Onnes Laboratory, Universiteit Leiden, 2300 RA Leiden, The Netherlands

Electrostatic charging of insulating fine particles can be responsible for numerous phenomena ranging from lightning in volcanic plumes to dust explosions. However, even basic aspects of how fine particles become charged are still unclear. Studying particle charging is challenging because it usually involves the complexities associated with many particle collisions. To address these issues we introduce a method based on acoustic levitation, which makes it possible to initiate sequences of repeated collisions of a single sub-millimeter particle with a flat plate, and to precisely measure the particle charge in-situ after each collision. We show that collisional charge transfer between insulators is dependent on the hydrophobicity of the contacting surfaces. We use glass, which we modify by attaching nonpolar molecules to the particle, the plate, or both. We find that hydrophilic surfaces develop significant positive charges after contacting hydrophobic surfaces. Moreover, we demonstrate that charging between a hydrophilic and a hydrophobic surface is suppressed in an acidic environment and enhanced in a basic one. Application of an electric field during each collision is found to modify the charge transfer, again depending on surface hydrophobicity. We discuss these results within the context of contact charging due to ion transfer and show that they lend strong support to OH$^-$ ions as the charge carriers.

# I. INTRODUCTION

Electrostatic charging of fine particles forms the basis for many applications, including electrophotography [1], powder coating [2], and air filtration [3]. One way of building up large amounts of charge in systems of small particles is through repeated collisions, a charging process that is thought to be responsible for phenomena such as lightning inside volcanic ash clouds [4], strong electric fields in sand storms [5] and dust devils [6], enhanced particle clustering [7, 8] or segregation [9] in granular materials and fluidized beds, and potentially catastrophic dust explosions [10]. Known parameters that affect particle charging include particle size [11-13], external electric field [14, 15], collisional energy [16], and ambient atmosphere [17, 18]. However, despite the importance of controlling phenomena due to collisional charging, the underlying mechanism is not well understood. Basic aspects are still being debated, such as how insulators, which have very low charge mobility, can transfer large amounts of charge when they are brought into contact [19]. Even the nature of the charge carrier associated with contact charging has not been settled, and several competing mechanisms have been proposed, including electron transfer [20, 21], ion transfer [22], and transfer of nanoscale pieces of charged material [23].

Interestingly, there appears to be a strong correlation between the hydrophilicity of insulators without ionic functional groups (nonionic insulators) and their charging behavior as characterized by the triboelectric series [22, 24-27]. Hydrophilic insulators such as glass are commonly closer to the top of the triboelectric series and charge positively after contacting other materials, while hydrophobic insulators such as Teflon are closer to the bottom of the triboelectric series, charging negatively. One interpretation has been that this is related to the Lewis acid-base properties of materials, where Lewis acids act as electron acceptors and Lewis bases as electron donors [25, 28, 29]. However, as McCarty and Whitesides point out [22], Lewis acid-base theory only considers electron sharing and cannot be used to distinguish between the transfer of electron or ions. A mechanism based on electrons also makes a correlation between hydrophobicity and charging behavior difficult to rationalize given that the ionization potentials for polymers do not follow their propensity of charging [30]. Furthermore, for contact charging of glassy dielectrics such as $ZrO_2$, electron transfer was ruled out recently by thermoluminescence measurements, which showed that the number of electronic trap states available is several orders of magnitude too small to account for the magnitude of charging observed [12].

Ion transfer as the cause for contact charging provides an attractive alternative to electron transfer. Insulators with ionic functional groups at their surface are known to develop charge of the opposite sign of the mobile ions during contact charging [31-35].

For nonionic insulators, it has been suggested that adsorbed water may be responsible [22]. Under ambient conditions, it only takes microseconds for water molecules to be adsorbed on any surface [36], and even under high vacuum it requires extensive baking to remove adsorbed water [37]. Water adsorbed from the ambient atmosphere has been found to enhance contact charging of insulators at relative humidity (RH) up to at least 40% [31, 38] (at high RH, contact charging can decrease due to higher surface conductivity, which allows charge to leak to ground [18, 31, 39]). Electrokinetic measurements and x-ray photoelectron spectroscopy revealed the accumulation of $OH^-$ ions at solid-water interfaces in hydrophobic insulators [40], hydrophilic self-assembled monolayers [41, 42], ice [43], and metals [44]. Recently, Burgo *et al.* demonstrated that water is positioned at the top of the triboelectric series [45]; that is, water develops positive charge after flowing over any other solid surface. Experiments on flow electrification of water by other research groups also showed the same direction of charge transfer [46-49]. All of these results can be ascribed to the adsorption of $OH^-$ ions at solid-water interfaces, which causes positive charging of water and negative charging of solids.

This points to a scenario whereby contact between two insulating solids could lead to a net charge transfer via $OH^-$ ions whenever there is an asymmetry in the coverage of absorbed water on the contacting surfaces. This will happen most likely when a hydrophilic and a hydrophobic surface come into contact. A particularly interesting consequence of these ideas is that they can provide new insights regarding the contact charging of objects made from the same material [11-13, 19, 20, 50], *i.e.*, in a situation where the triboelectric series has no predictive power.

Here we test this scenario for collisional contact charging of fine particles, where systematic, quantitative experiments have been notoriously difficult. To date, the most common technique for studying the charging of fine particles uses Faraday cup devices [9, 51, 52]. These devices involve large numbers of particles, which, as a result of multiple, uncontrolled particle-particle and particle-wall collisions, hinders the interpretation of the results. While there are experimental techniques for making precise measurements of the impact charging of a single sub-millimeter particle [53-56], what has been lacking is a method for tracking the evolution of the charging process over repeated, highly controlled collisions.

In this paper, we introduce such method by combining high-speed videography and acoustic levitation [57-61]—a contact-free and material-independent technique of object manipulation. This enables us to create automated, computer-controlled collision sequences followed by *in situ* charge measurement after each collision. As a result, we can narrow down the complex problem of many-particle collisional charging to following repeated collisions of a single particle with a target plate. In our experiments we change the hydrophobicity of the particle and plate surfaces, vary the ambient gas environment, and apply external electric fields during particle-plate collisions. Our findings provide strong evidence that hydrophilicity plays a key role in contact charging and they offer new insights into how contact charging of fine particles can be controlled.

## II. EXPERIMENTAL SETUP AND PROCEDURE

The basic idea for our setup is to create with ultrasound a switchable particle trap and to extract the particle's charge from its resonant oscillatory motion inside that trap. In a typical measurement sequence, the particle is initially suspended above a flat plate by acoustic levitation. Next, an oscillating electric field is applied and the particle charge $q$ is extracted from the response to the field. To initiate a collision, the acoustic trap is then briefly switched off. This causes the particle to drop due to gravity, collide with the target plate below, and bounce up. By appropriately timing when the acoustic field is turned back on (~12 ms in our experiments) the particle becomes trapped again, ensuring that there was precisely one collision per drop (see SI movie). After each drop, the AC electric field is applied to extract $q$. This sequence is repeated under computer control over 100s or 1,000s of collision events, enabling precise tracking of collisional charge transfer.

The experimental setup is shown in Fig. 1(a). A function generator and a high voltage amplifier provided the driving voltage signal for a transducer (Hesentec Rank E) that generated ultrasound at 146.5 kHz. A flat target plate underneath the transducer acted as a reflector. Setting the distance between the transducer and the sound-reflecting target plate to half the wavelength of sound ($\lambda/2$) created a standing wave with a single pressure node at $\lambda/4$, in which a sub-millimeter particle could be trapped. To also confine the particle horizontally, we machined a small cylindrical depression (800 μm diameter, 125 μm deep) into the flat surface of the transducer and adjusted the distance between the bottom of the depression and the target plate to match $\lambda/2$ ($\approx$ 1.2 mm).

For a spherical particle with radius $R$ much smaller than $\lambda$ (in our experiments $R \approx$ 100 μm and thus $R/\lambda < 0.1$), the acoustic potential $U$ derived by Gor'kov is [62]

$$U = 2\pi R^3 \left(\frac{\langle p^2 \rangle}{3\rho c^2} - \frac{\rho \langle v^2 \rangle}{2}\right), \tag{1}$$

where $\rho$ is the density of the ambient gas, $c$ is the speed of sound, and $\langle p^2 \rangle$ and $\langle v^2 \rangle$ are the mean square amplitudes of the acoustic pressure and velocity of the gas. To obtain expressions for the latter two, we start from the acoustic velocity potential for a standing sound wave along the vertical axis ($y$-axis), given by [63]

$$\phi = -\frac{v_0}{k} \cos ky \sin \omega t, \tag{2}$$

where $v_0$ is the maximum acoustic velocity, $k = 2\pi/\lambda$ is the wave number, and $\omega = kc$ is the angular frequency of sound. This leads to acoustic pressure and velocity for the standing sound wave along the y-axis according to

$$p = -\rho \frac{\partial \phi}{\partial x} = \rho c v_0 \cos ky \cos \omega t \qquad (3)$$

and

$$v = \frac{\partial \phi}{\partial y} = v_0 \sin ky \sin \omega t. \qquad (4)$$

Inserting Eqs. (3) and (4) into (1) we obtain $U$. Figure 1(b) shows the dimensionless acoustic potential $\widetilde{U} = U/(\pi R^3 \rho v_0^2)$, simulated with COMSOL Multiphysics 5.2. The particle can be trapped at the vertical equilibrium point, where gravity is balanced by the acoustic radiation force

$$F_{ay} = -\frac{\partial U}{\partial y} = \frac{5}{6}\pi R^3 \rho v_0^2 k \sin 2ky. \qquad (5)$$

To obtain the net charge $q$ of a levitated particle, we applied a vertical electric field across the gap between an aluminum block underneath the target plate and the grounded transducer (Fig. 1(a)). Via a computer-controlled relay, a large DC field was applied to extract the particle's charge polarity by observing its up or down movement with the camera, and an AC field was applied (using a signal generator and an amplifier) to drive the particle into resonance. From the measured resonance frequency and amplitude at resonance we then extracted the charge magnitude $|q|$ by a numerical procedure, as described next.

The equation of motion of the particle along the y-axis can be expressed as

$$m\ddot{y} + 2m\beta\dot{y} = F_{ay} - mg + |q|E_0 \sin(\omega_E(t)t), \qquad (6)$$

where $m$ is the particle mass, $2m\beta\dot{y}$ is the air drag on the particle, $mg$ is the force of gravity, and $E_0$ and $\omega_E$ are the amplitude and angular frequency of the applied AC electric field.

For a given particle, we computed $m$ from the material density and the particle size as measured with an optical microscope (Fig. 1(c)). The parameter $\beta$ was obtained from fitting the envelope of the free vertical oscillation of the particle after it was picked up from the target plate by applying a strong acoustic radiation force in the absence of any E-field (Fig. 1(d), $\beta \approx 2.4$ s$^{-1}$ for the example shown). The field strength $E_0$ was obtained from modeling with COMSOL the electric field between the aluminum block and the transducer, including the small depression in the transducer surface and the presence of the glass target plate. $F_{ay}$ is provided by Eq. (5) and can be expressed as

$$F_{ay} = \frac{5}{6}\pi R^3 \rho v_0^2 k \sin\left(\frac{4\pi y}{\lambda}\right) \equiv F_{a0} \sin\left(\frac{4\pi y}{\lambda}\right). \qquad (7)$$

As a function of the parameters $|q|$ and $F_{a0}$, Eq. (6) can be solved numerically to obtain simulated particle trajectories $y_s(t)$. Each such trajectory is characterized by two features, a resonance frequency and an amplitude at resonance, that can directly be compared to experiment. We then treated $|q|$ and $F_{a0}$ as variables and used a Nelder–Mead simplex algorithm [64] to minimize the squared relative error

$$\delta = \left(\frac{A'_{max}-A_{max}}{A_{max}}\right)^2 + \left(\frac{f'_{max}-f_{max}}{f_{max}}\right)^2 \tag{8}$$

between simulated (primed) and experimentally measured (unprimed) amplitudes and frequencies.

To find these amplitudes and frequencies, a linear frequency sweep was applied, starting at some frequency $f_i$. This means $\omega_E(t) = 2\pi(\alpha t/2 + f_i)$, where $\alpha$ is the sweep rate. From the experimentally measured trajectories and their simulated counterparts the resonance frequencies and amplitudes were then extracted via a Hilbert transform [65] (see Appendix B). The minimization was terminated when $\delta$ became less than $10^{-7}$. This provided a well-defined, automated path that led from experimental trajectory data to the charge magnitude $|q|$.

The close agreement between experimental and simulated data is shown in Figs.1(e)&(f). The figures give an example of $y(t)$ in response to an AC electric field that was frequency-swept from 50 Hz to 150 Hz in 15 seconds. The particle exhibited a resonance with $A_{max} \approx 52$ μm at $f_{max} \approx 93$ Hz.

One particular advantage of our method is that by including a time-dependent $\omega_E$ in Eq. (6), experimental frequency sweeps can be fast and $|q|$ can still be extracted reliably, even if in those cases the apparent resonance frequency and associated maximum amplitude have shifted because the particle did not have sufficient time to equilibrate. This is shown in Fig. 1(g).

Based on Eq. (6), $|q|$ scales as $m/E_0$. As a result, the uncertainty $\Delta|q|$ of the measured charge magnitude becomes smaller with larger $E_0$. Our experiments used AC electric fields in the range $3.3 \times 10^3$ V/m $< E_0 < 1.6 \times 10^5$ V/m, giving a charge uncertainty $1 \times 10^3$ e $\lesssim \Delta|q| \lesssim 6.5 \times 10^4$ e, in units of the elementary charge e = $1.6 \times 10^{-19}$ C. Our detection limit for the particle charge arises from the difficulty of measuring $A_{max}$ and $f_{max}$ when the particle oscillation amplitude falls below ~3 μm.

All experiments were conducted at 40% RH inside a Plexiglas chamber filled with an adjustable mixture of dry nitrogen gas and nitrogen gas saturated with water vapor (unless otherwise stated). The levitated particle was imaged from the side by a high-speed video camera (Phantom v12; 1000 frames per second) using backlight illumination. The particle position was tracked with the algorithm developed by Crocker and Grier [66].

The measurements reported here focus on the specific case were both the particle (Fig. 1(c)) and the target plate were borosilicate glass. This material is naturally hydrophilic due to polar -OH groups at its surface. This is shown by contact angle measurements (Fig. 2(a)). To investigate the effect of hydrophobicity on particle charging, we made glass particles or glass plates hydrophobic by attaching nonpolar methyl ($-CH_3$) groups via silanization (see Appendix A for details on sample preparation).

## III. RESULTS AND DISCUSSION
### A. Charging of hydrophilic and hydrophobic particles

In Fig. 2(b), we plot the evolution of the net charge $q$ on a levitated particle, after it has collided with the target plate $N$ times. For a hydrophilic glass particle colliding with a hydrophobic glass plate (red trace), we find that, for more than 500 successive collisions, the particle steadily accumulates positive charge at a constant rate $dq/dN \approx +6600$ e/collision. If the conditions are inverted and a hydrophobic glass particle collides with a hydrophilic plate (blue trace), we find that the particle accumulates negative charge at a rate $dq/dN \approx -6400$ e/collision, *i.e.*, the amount of charge transferred is similar but in the opposite direction. For contact between similar surfaces (hydrophilic particle vs. hydrophilic plate and hydrophobic particle vs. hydrophobic plate), charging is very much suppressed (green and black traces). In these experiments the number of collisions $N$ was limited by the charge magnitude: when the charge reached $|q| \approx 4 \times 10^6$ e, the attractive electrostatic force between the particle and the target plate became comparable to the acoustic lifting force, and after colliding the particle stuck to the plate.

Taken together, these data show clearly that collisional contacts drive hydrophilic glass surfaces positive and their hydrophobic counterparts negative. To explain our results, we start from the suggestion by McCarty and Whitesides that contact charging between nonionic insulators very generally might be due to the transfer of $OH^-$ ions from molecularly thin layers of water adsorbed on the surfaces at ambient conditions [22]. An important point is that such layers do not form a continuous film but instead break up into islands or patches, even on a hydrophilic material like atomically flat mica [67]. At 40% RH the surface coverage with water was found to be only ~50% on mica, with water patches ranging in spatial extent from tens of nanometers to over a micron, and even at 90% RH incomplete coverage was observed. A second important point is that negative charge accumulates at the interface of a water layer with a hydrophobic surface. While many details of the configuration of aqueous ions at surfaces are still not fully understood and various mechanisms supporting the observed negative charge are under debate [68-70], molecular-dynamics simulations have shown that $OH^-$ ions tend to be more localized at the interface with a hydrophobic surface, while $H^+$ ions are more mobile and can thus diffuse more readily away [71]. Alternatively, an excess of $OH^-$ ions at such interface has been ascribed to enhanced autolysis of interfacial water molecules [69, 72, 73].

Based on this propensity for $OH^-$ ions to accumulate at hydrophobic interfaces, we suggest a possible scenario for the charge transfer between nonionic insulators with different hydrophobicity and thus different amounts of adsorbed water. This is sketched in Fig. 2(c). When an adsorbed water island (wet patch) on one surface contacts an area without water molecules (dry patch) on another surface, OH⁻ ions from the wet patch preferentially accumulate at the interface between water and dry patch. When the contact is broken (*e.g.*, at the end of a collision), the greater affinity of OH⁻ for the interface together with the larger mobility of H⁺ then implies that there is a high likelihood that OH⁻ is left behind (together with some neutral water). This leads to negative charge transfer from wet to dry patch after separation.

This scenario is consistent with recent experiments, which found that solid materials tend to become negatively charged after contacting water [45]. For hydrophilic insulators, water molecules are adsorbed from the atmospheric moisture due to the formation of hydrogen bonds [74]. Although hydrophobic insulators can still adsorb water due to ubiquitous dispersion forces [74, 75], these forces are smaller than hydrogen bonds, which can lead to larger dry patches on a hydrophobic surface. A hydrophilic surface has the larger amount of adsorbed water and therefore, after separation, it leaves a net negative charge on the hydrophobic surface. Another factor may be that a hydrophobic insulator can have a higher affinity for OH⁻ ions, leading to more OH⁻ adsorption on the hydrophobic surface. This is based on electrokinetic studies, which find that the zeta potential is generally more negative with increasing hydrophobicity [43, 76, 77].

Our scenario implies that the net charge transfer is determined by the difference in coverage of water islands between the two contacting surfaces and by the affinities of OH⁻ and H⁺ ions for different materials. An immediate consequence is that two contacting surfaces of similar hydrophobicity or hydrophilicity, with their statistically similar coverage of water patches, should exhibit a greatly reduced net charge transfer, as already seen in Fig. 2(b).

In fact, at these low charging rates, subtle differences in material fabrication, preparation or handling become apparent. Figure 2(d) shows different trials of charge transfer measurements for a hydrophilic glass particle colliding with a hydrophilic glass plate. We observe both directions of charge transfer, at charging rate from $dq/dN = -460$ to 900 e/collision. As expected, the charging rate is even smaller for a hydrophobic particle colliding with a hydrophobic glass plate, with $dq/dN$ from −26 to 80 e/collision (Fig. 2(e)), because fewer wet patches, and thus fewer transferable ions, are available.

*In situ* contact charging measurements by other research groups imply that the surface charge should eventually saturate with $N$ once the transferable ions have been depleted, or should exhibit rapid discharge events when the charge per unit surface area has become sufficiently large [33, 50, 78]. By contrast, the results in Fig. 2 exhibit no sign of charge saturation up to $|q| = 4 \times 10^6$ e. This is explained by the fact that the particles rotate while trapped and oscillated [79]. Therefore, different spots of the particle contact the target plate during successive collisions. With a maximum contact area $A_c$ during particle collisions of ~20 μm$^2$ in our experiments (Appendix C) and a total particle surface area of ~$1.3 \times 10^5$ μm$^2$, we estimate that it would take, on average, over 7,000 collisions until the same patch is involved in a collision for the second time. The highly linear $q(N)$ dependence we found in all cases thus demonstrates that any wet and dry patches must have been much smaller than 20 μm$^2$, so that the particle surfaces appeared homogeneous when averaged over the contact area.

### B. Particle charging in acidic/basic atmospheres

If the contact charging is indeed due to OH$^-$ ions, then changing the chemical composition of the ambient atmosphere, which will alter the number of available OH$^-$ ions in the adsorbed water layers, should influence the amount of charge transferred per collision. To test this, we changed the ambient atmosphere by introducing acidic or basic vapors to the experimental chamber, similar to Ref. [38]. These vapors were prepared by bubbling nitrogen gas into 1M aqueous solutions of acetic acid or ammonia, respectively. The chamber atmosphere was stabilized for two hours before each experiment.

In Fig. 3, we plot the charging rate of a hydrophobic particle colliding with a hydrophilic glass target plate as a function of relative humidity in acidic, basic, and neutral environments. Over the relative humidity range studied (10-50% RH), the negative charging rate of the particle in a neutral atmosphere (Fig. 2(b)) is further enhanced in a basic atmosphere, while it is suppressed in an acidic atmosphere. This makes sense in light of the electrokinetic studies, which showed that OH$^-$ and H$^+$ ions have a much stronger affinity for hydrophobic surfaces than other ions [40, 41, 43, 76], whereas the acetate and ammonium counter-ions behave indifferent [80-82]. Thus, the enhancement and suppression of the charging rate in basic and acidic environments can be attributed to different numbers of OH$^-$ ions available in the adsorbed water, which are the ions predominantly transferred from a hydrophilic to a hydrophobic surface during contact.

Our experiments could only reach RH values around 50%. Controlled levitation-collision sequences become difficult to perform with larger humidity due to the increasingly strong capillary forces between particle and target plate. While the data in Fig. 3 stays relatively flat up to 50% RH, larger humidity eventually will let charge leak away by increasing the conductivity. Previous studies showed, however, that charging of hydrophobic particles does not decrease until 60 % RH is reached [18, 39].

### C. Particle charging controlled by electric fields

Particle charging induced by electric fields has been observed previously [14, 15],

but the *kinetics* of how a single particle gets charged due to repeated collisions in an electric field has not been demonstrated. In our setup we can explore this straightforwardly by applying a DC electric field during particle collisions. In Fig. 4(a) we plot $q(N)$ for a hydrophilic glass particle colliding with a hydrophilic glass target plate in a field $E = \pm\, 8 \times 10^5$ V/m pointing up (to the particle) or down (to the target plate). In this particular example, the particle started out negatively charged and became less negative when colliding in an electric field pointing up. Then $q$ went through zero and became more positive, charging at the same rate. As this demonstrates, we found that the direction of charge transfer for a particle was not affected by its own polarity. Yet, sharp transitions of charging rate were observed after changing the direction of the electric field. The particle charged negatively when the electric field pointed down, and it charged positively when the electric field again pointed up.

In these experiments the acceleration of a particle in the electric field was less than ~ 0.7 m/s$^2$ and measurements were stopped before the acceleration became larger. At this *maximum* acceleration value the $E$-field caused no more than a ~3% change in the impact velocity ($V_{imp} = (2gh)^{1/2} \approx 0.11$ m/s without electric field applied) and a ~3% change in the maximum contact area during collision $A_c$.

We can explain the observed field-induced charging by considering the migration of ions due to the external electric field when, during particle collisions, the two contacting surfaces are brought into contact. For contact between two wet patches, no net charge transfer happens without a field; however, with an applied field, ions of different polarity migrate across the interface in different directions, inducing a field-controlled net transfer of charge after separation, as illustrated in Fig. 4(b)-(e). Similarly, when a wet and a dry patch collide (see Fig. 2(c)), the applied field drives OH$^-$ ions in the wet patch either towards or away from the interface with the dry patch and thus controls the amount of charge transfer. For the contact between two dry patches, no ions will be transferred.

As before when $E = 0$ during collisions, no obvious sign of charging saturation is seen in Fig. 4(a). This again is because of particle rotation, so that, at least up to the collision numbers $N$ tracked in our experiments, different spots of the particle hit the target plate. Therefore, for each collision new patches are brought into contact and contribute to the field-induced charge transfer. In addition, if we assume uniform charge distribution on the particle, the electric field $(1/4\pi\varepsilon_0)(q/R^2)$ due to the charge buildup is no larger than $8 \times 10^3$ V/m for the data in Fig. 4(a). This is two orders of magnitude smaller than the external applied electric field $E = 8 \times 10^5$ V/m. Therefore, the built-up particle charge is unlikely to affect the collisional charging, and the external electric field still dominates the charge transfer.

To isolate the field-induced contribution to the total charging rate we subtract off the charging rate measured in the absence of the field, which can be zero, positive, or negative (Fig. 2(b)) and can even depend on subtle differences in the average number of wet and dry patches (Fig. 2(d)). Figure 4(f) shows $(dq/dN)_E − (dq/dN)_0$ versus $E$ for a hydrophilic glass particle colliding with a hydrophilic glass target plate, where $(dq/dN)_E$ is the rate for a particle colliding in a DC field, and $(dq/dN)_0$ is the rate for the same particle and target plate without the field. A linear relationship between is observed, with slope $1.25 \times 10^{-3} \pm 8 \times 10^{-5}$ (e/collision)/(V/m).

Along the same lines we expect significantly reduced coverage for a hydrophobic glass particle colliding with a hydrophobic glass target plate in a DC electric field. Indeed, the slope of the linear fit of $(dq/dN)_E − (dq/dN)_0$ vs. $E$ is only $2.8 \times 10^{-5} \pm 3 \times 10^{-6}$ (e/collision)/(V/m) (Fig. 4(g)), a reduction of nearly two orders of magnitude compared to the case of hydrophilic glass surfaces.

## IV. CONCLUSIONS

We showed for the case of glass particles colliding with a glass plate that contact charging is controlled by the hydrophobicity of the contacting surfaces: hydrophilic surfaces accumulate positive charge, whereas hydrophobic surfaces obtain negative charge. Contact charging between a hydrophilic and a hydrophobic surface is suppressed (enhanced) in an acidic (basic) atmosphere. We further showed that the application of an external electric field during particle collisions can be used to control the direction of charge transfer. These findings provide strong evidence in support of a contact charging mechanism for nonionic insulators that relies on the transfer of $OH^−$ ions. Presumably these ions are transferred from molecularly thin patches of water on one of the colliding surfaces to a dry patch on the other. In this scenario, hydrophilic surfaces simply contain more wet patches than hydrophobic surfaces do, thereby increasing the likelihood of donating $OH^−$ ions. This also can explain why $E$-field-induced charging is significantly larger for collisions between hydrophilic surfaces than between hydrophobic ones. Furthermore, this scenario provides an explanation for the collisional charging of insulating particles comprised of the very same material, where there are too few electrons available to act as the charge carriers. Adsorbed water, by contrast, provides a plentiful source of $OH^−$ ions, which means that larger particles can donate more $OH^−$ than smaller particles and thus tend to charge positively.

The acoustic-levitation-based technique demonstrated here can be extended straightforwardly to other particle types and sizes as well as particle-plate material combinations. The only significant limitation is that the particle should be shaped such that after each collision it bounces upward vertically and can be recaptured by the acoustic field. In principle, it should also be possible to vary the collision velocity by modulating the acoustic field or by accelerating (or decelerating) the particle with an appropriately chosen E-field pulse prior to each collision.

**Acknowledgements**

We would like to thank Thiago Burgo, Troy Shinbrot, Andrei Tokmakoff, and Tom Witten for insightful discussions. This work was supported by the National Science Foundation under grant DMR-1309611. N.J. acknowledges support by the Chicago MRSEC through NSF DMR-1420709, which also provided the high-speed camera through its shared experimental facilities.

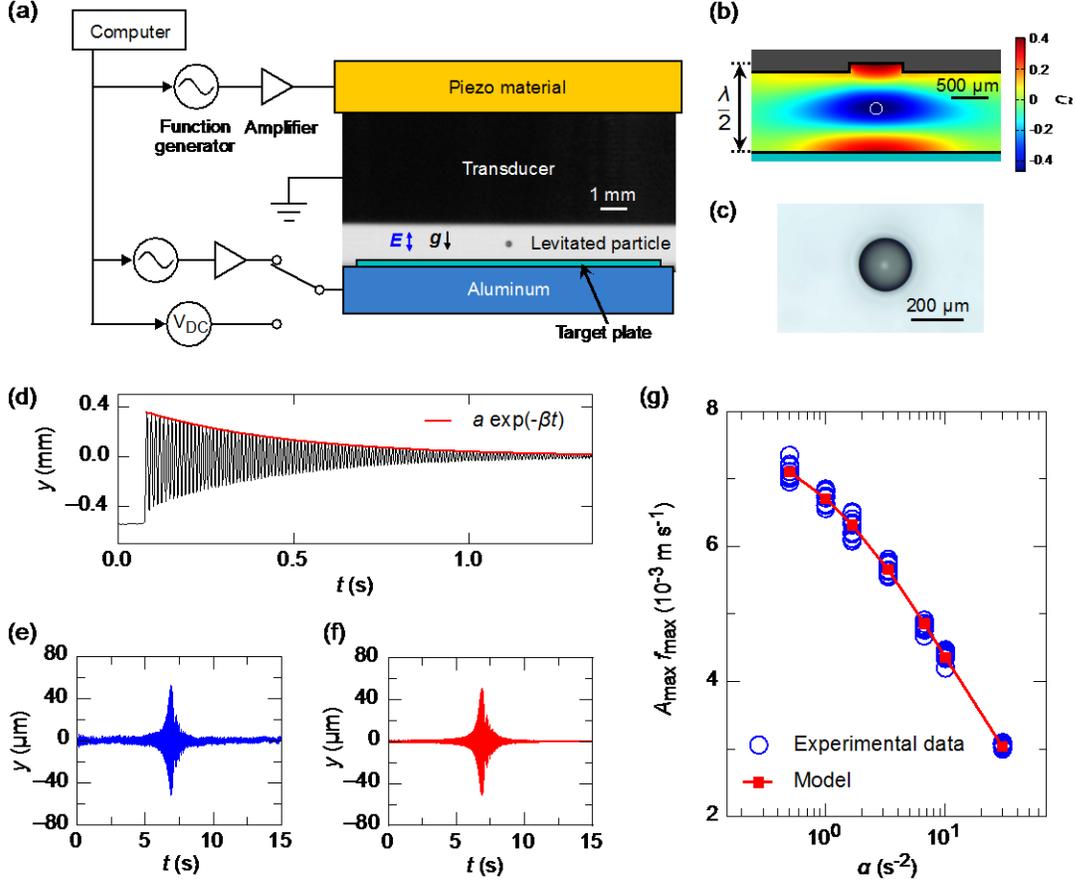

Fig. 1. Particle charge measurement with acoustic levitation. (a) Experimental setup. A sub-millimeter particle is acoustically levitated between a grounded ultrasonic transducer and a sound-reflecting target plate. A piece of aluminum underneath the target plate is connected to an AC or DC voltage source. The particle was backlit and filmed from the side with a high-speed camera. The whole setup is enclosed in a chamber to control the ambient gas. (b) Simulated dimensionless acoustic potential $\tilde{U}$. The blue central region forms the trap. The white circle gives the approximate size and position of a levitated particle. (c) Optical microscope image of a glass particle. (d) Free oscillation of a particle lifted from the plate and trapped by suddenly turning on the acoustic field. The red line is a fit to an exponential decay due to air drag. (e)&(f) Experimental data (e) and model (f) for the position $y(t)$ of a trapped particle in a frequency-swept AC electric field. (g) Comparison between experimental data and model for the product of maximum oscillation amplitude $A_{max}$ and resonance frequency $f_{max}$ as a function of frequency sweep rate $\alpha$.

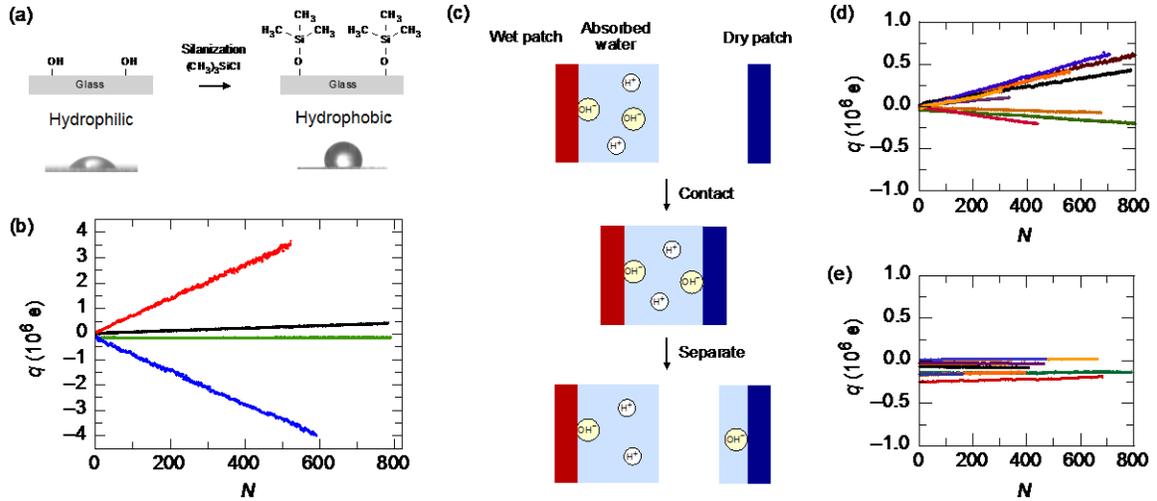

Fig. 2. Charge transfer between glass surfaces. (a) Glass surface silanized with trimethylchlorosilane $(CH_3)_3SiCl$. The contact angles between a water drop and the glass plate before and after silanization were $50 \pm 3°$ and $110 \pm 3°$, respectively. (b) Examples of the evolution of particle charge $q$ with the number of particle-plate collisions $N$ for a hydrophilic particle against a hydrophobic plate (red), hydrophobic particle against hydrophilic plate (blue), hydrophilic particle against hydrophilic plate (black), and hydrophobic particle against hydrophobic plate (green). In the experiments corresponding to the red and blue curves, we used a lower amplitude of the AC electric field ($3.3 \times 10^3$ V/m) to prevent highly charged particles from colliding with the transducer or target plate during the field-driven oscillations, which resulted in a higher statistical error of the charge measurements. (c) Proposed mechanism of charge transfer between a wet patch and a dry patch on two contacting surfaces. (d)&(e) Different trials showing $q(N)$ for a hydrophilic particle against a hydrophilic plate (d) and a hydrophobic particle against a hydrophobic plate (e).
14

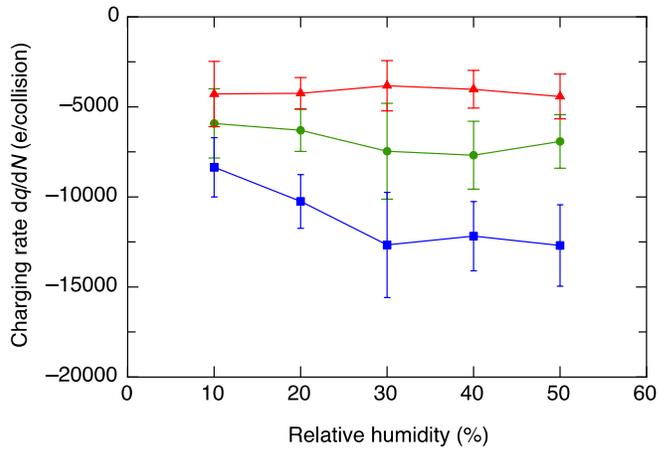

Fig. 3. Charging rate d*q*/d*N* for a hydrophobic glass particle colliding with a hydrophilic glass target plate as a function of relative humidity in acidic (red), neutral (green), and basic (blue) atmospheres.

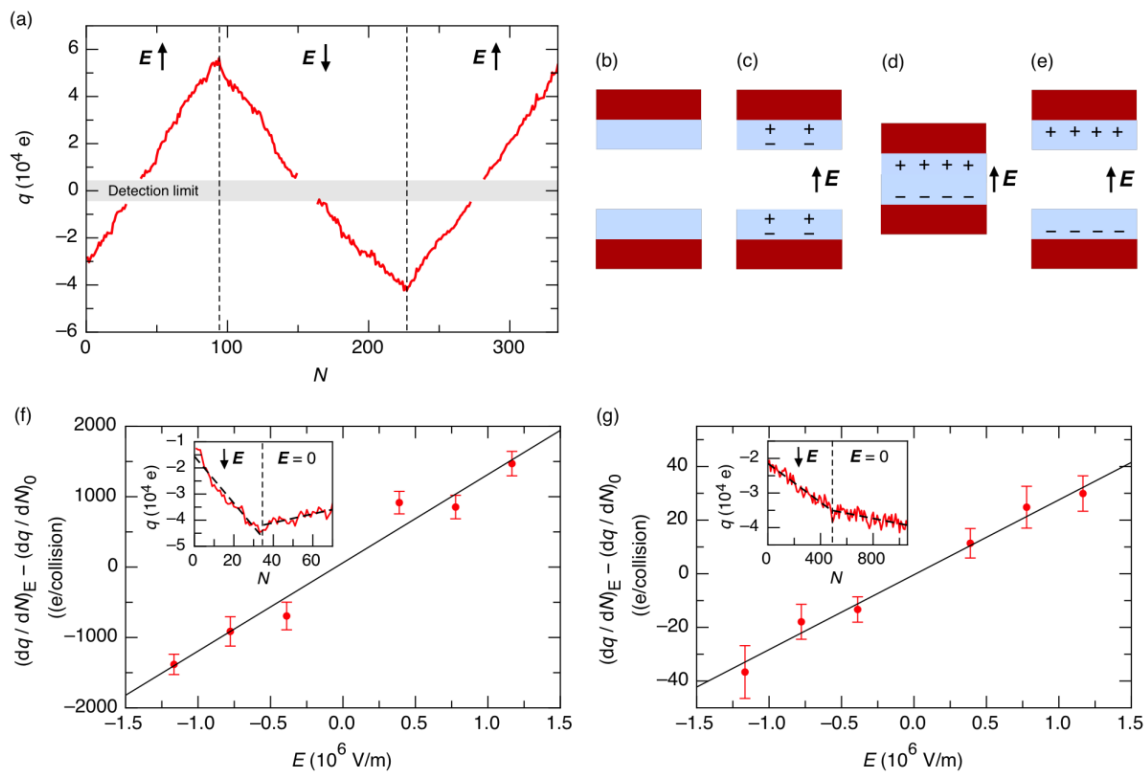

Fig. 4. Field-induced contact charging. (a). Charge $q$ on a hydrophilic glass particle colliding with a hydrophilic glass target plate in an electric field pointing up or down ($E = \pm 8 \times 10^5$ V/m) as a function of number of collisions $N$. (b)-(e) Charging mechanism of two surfaces contacting in an electric field. Here we only plot the contacts where no dry area is involved on either side. (f)&(g) Charging rate in an electric field minus the charging rate without electric field as a function of field strength $E$, plotted for a hydrophilic glass particle colliding with a hydrophilic glass target plate (f) and for a hydrophobic glass particle colliding with a hydrophobic glass target plate (g). Insets show representative examples of the corresponding $q(N)$ data before subtraction, in both cases for $E = -8 \times 10^5$ V/m.

**Appendix A: Sample preparation**

Borosilicate glass particles (Cospheric; density $\rho_p = 2200$ kg/m$^3$, diameter $D = 204 \pm 8$ μm) and borosilicate glass target plates (Fisher Scientific, Catalog no. 12-542B cover slides; 22 mm × 22 mm, 0.15 ± 0.02 mm thick) were cleaned in an ultrasonic bath of acetone, ethanol, and DI water for one hour each, and dried in a vacuum chamber with an oil-free pump. Hydrophobic glass particles and target plates were prepared by silanizing hydrophilic substrates, adapting the method used by Burgo *et al.* [83]: glass particles and cover slides were boiled in anhydrous ethanol for 2 hours and then immersed in a base bath for 1.5-3 hours. The glass substrates were then rinsed with water and ethanol and dried for 12 hours at 65°C. Then, they were then immersed in a 10% (w/w) solution of chlorotrimethylsilane (Sigma-Aldrich) in ethanol for 8 hours, washed with ethanol, dried at 70°C on a hot plate, and stored in a desiccator.

**Appendix B: Using Hilbert transform to obtain the maximum amplitude of oscillation and the resonance frequency**

We can obtain the analytic signal $Y(t)$ for the vertical position $y(t)$ of the levitated particle
$$Y(t) = y(t) + i\tilde{y}(t) = A(t)e^{i\varphi(t)}, \tag{9}$$
where $\tilde{y}(t)$ is the Hilbert transform of $y(t)$ defined by [65]
$$\tilde{y}(t) = \hat{H}[y(t)] = \frac{1}{\pi}\int_{-\infty}^{\infty} \frac{y(\tau)}{t-\tau} d\tau. \tag{10}$$
The instantaneous amplitude $A(t)$ and frequency $f(t)$ of $y(t)$ can be obtained by
$$A(t) = |Y(t)| = \sqrt{y^2(t) + \tilde{y}^2(t)} \tag{11}$$
and
$$f(t) = \frac{1}{2\pi}\frac{d\varphi(t)}{dt}, \tag{12}$$
where $\varphi(t)$ is the instantaneous phase angle, given by
$$\varphi(t) = \arctan\left(\frac{\tilde{y}(t)}{y(t)}\right). \tag{13}$$
In Fig. 5, we plot the zoomed-in data of $y(t)$ in Fig. 1(e) and its corresponding $A(t)$ and $f(t)$. We define the maximum amplitude of oscillation $A_{\max}$ to be the maximum point of $A(t)$ at $t = t'$, and the resonance frequency $f_{\max}$ to be its corresponding instantaneous frequency ($f_{\max} = f(t')$).

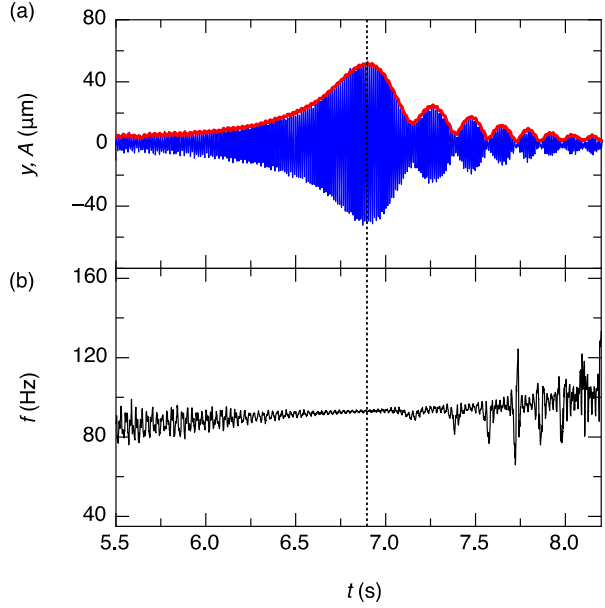

Fig. 5. Instantaneous amplitude and frequency obtained from the Hilbert transform. (a) Zoomed-in data of $y(t)$ from Fig. 1(e) (blue) and its instantaneous amplitude $A(t)$ (red). (b) Instantaneous frequency $f(t)$ of $y(t)$ in (a). The vertical dotted line indicates the time $t'$ when $A = A_{max}$ and $f = f_{max}$.

**Appendix C: Estimation of maximum contact area during particle-plate collisions**

The maximum contact area during collision $A_c$ due to elastic-plastic deformation for a particle being dropped onto a flat surface can be estimated with the Hertz contact theory [84, 85]:

$$A_c = 0.41 \pi D^2 \sqrt{\frac{\rho_p}{Y_p}} \left( V_{imp} - 0.05 \sqrt{\frac{\kappa^4 Y_p^5}{\rho_p}} \right), \qquad (14)$$

where $D = 204 \pm 8$ μm, $\rho_p = 2200$ kg/m$^3$ and $Y_p = 190$ MPa are diameter, density and yield pressure of the particle, respectively, supplied by the manufacturer, $V_{imp} = (2gh)^{1/2} = 0.11$ m/s is the impact velocity, $h = 0.6$ mm is the drop height, $\kappa = (1 - v_1)^2/E_1 + (1 - v_2)^2/E_2$ is the elasticity parameter, $v_1 = v_2 = 0.2$ and $E_1 = E_2 = 67.5$ GPa are Poisson's ratio and Young's modulus of borosilicate glass [86], and subscripts 1 and 2 denote the particle and the target plate, respectively. We obtain $A_c \sim 20$ μm$^2$ for a glass particle in our setup.

**Appendix D: Charges of levitated particles without any collisions**

To see the effect of the ambient N$_2$ gas on the charge $q$ of the particles, we levitated hydrophilic and hydrophobic glass particles and repeatedly measured $q$ without any collisions for ~25 hours. An example is shown in Fig. 6. This is much larger than the amount of time spent in our experiments of repeated particle collisions (typically less than 6 hours). No detectable charge transfer between the ambient gas and the levitated particles was found.

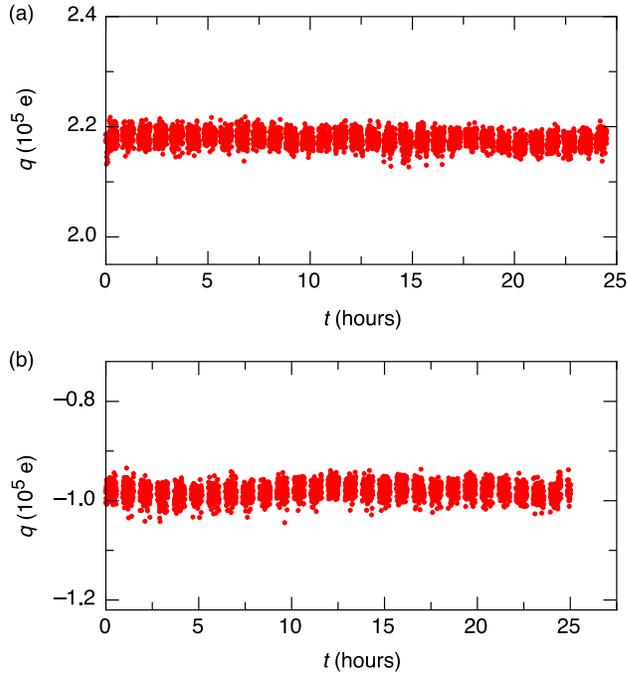

Fig. 6. Examples of charge $q$ on (a) a hydrophilic glass particle and (b) a hydrophobic glass particle levitated in $N_2$ at 40% RH vs. time $t$. The gaps along the time axis are due to halting high-speed video image acquisition while saving the buffered video to disk.